# Understanding Team Collaboration in Artificial Intelligence from the perspective of Geographic Distance

Xuli Tang[1[0000-0002-1656-3014]], Xin Li[1[0000-0002-8169-6059]], Ying Ding[2[0000-0003-2567-2009]] and Feicheng Ma[1[0000-0003-0187-0131]]

[1] School of Information Management, Wuhan University, Wuhan, China
[2] School of Information, University of Texas at Austin, TX, USA
`xulitang@whu.edu.cn, lucian@whu.edu.cn,`
`ying.ding@austin.utexas.edu, fchma@whu.edu.cn`

**Abstract.** This paper analyzes team collaboration in the field of Artificial Intelligence (AI) from the perspective of geographic distance. We obtained 1,584,175 AI related publications during 1950-2019 from the Microsoft Academic Graph. Three latitude-and-longitude-based indicators were employed to quantify the geographic distance of collaborations in AI over time at domestic and international levels. The results show team collaborations in AI has been more popular in the field over time with around 42,000 (38.4%) multiple-affiliation AI publications in 2019. The changes in geographic distances of team collaborations indicate the increase of breadth and density for both domestic and international collaborations in AI over time. In addition, the United States produced the largest number of single-country and internationally collaborated AI publications, and China has played an important role in international collaborations in AI after 2010.

## 1    Introduction

Team collaboration is defined as the process where researchers from various affiliations working together for common goals by sharing knowledges, resources and experiences [1]. It has recently become imperative in the field of artificial intelligence (AI) for serving humanity in more complicate situations. Take for an example, with the global outbreak of COVID-19, AI scientists have collaborated with virologists, clinicians, and epidemiologists worldwide [2], on automatic image diagnosis [3], drug repurposing [4], and global epidemic prediction [5].Team collaboration not only helps AI scientists deeply understand outputs of  AI algorithms, but also provides domain experts with insights on virus, which can speed up the process of combating COVID-19. However, both domestic and international collaboration in AI has not been prevalent yet [6]. Thus, it is worthwhile to understanding the status of team collaborations in AI and to identify the factors affecting collaborations in AI.

Team collaboration has been discussed in many disciplines, such as ecology [7] indigenous knowledge [8] and Zika virus [9]. Geographic distance is widely considered as one of the major factors affecting team collaboration. Parreiria et al. [7] found that 10% of team collaborations among countries can be explained by geographic distance and socioeconomic factors. Sidone et al. [10] concluded that the geographic proximity is beneficial for team collaboration and knowledge exchange. Few studies have explored how AI scientists collaborated with each other.

In this preliminary study, we aim to understand team collaboration in AI from the perspective of geographic distance. In past studies, the geolocations of countries' capitals are commonly adopted to calculate the distance among collaborators [11]. Thanks to the geographic data in MAG, we here employed the latitude and longitude of affiliations to precisely quantify the maximum, minimum and average geographic distances for team collaborations in AI. We conducted the geographic distance analysis of team collaboration in AI at domestic and international levels.

## 2    Methodology

### 2.1    Data and Processing

The data set used in this study has been derived from the Microsoft Academic Graph. We use publications in the subfields of artificial intelligence, machine learning, computer vision, nature language processing and pattern recognition to represent AI publications [12-13]. Bibliographic information of each publication such as title, year and abstract were extracted and stored in a local MySQL database. We recognized the country for each affiliation using its latitude and longitude. After removing publications without any affiliation information, we obtained 1,584,175 AI publications in1950-2019 with 4,998,781 unique authors belonging to 13,807 unique affiliations. 24.7 % of AI publications are multiple-affiliation publications, which are comprised of 177,794 single-country AI publications and 213,011 multiple-country AI publications.

Fig. 1 (a) shows the publication number distribution (blue) and the affiliation number distribution(orange) over years, in which the past 70 years has witnessed the exponentially increasing for both publications and affiliations. Fig. 1 (b) displays the relationship between the number of publications and the number of affiliations, which shows an approximate power law distribution. It indicates that most of AI publications were produced by a small number of highly productive affiliations.

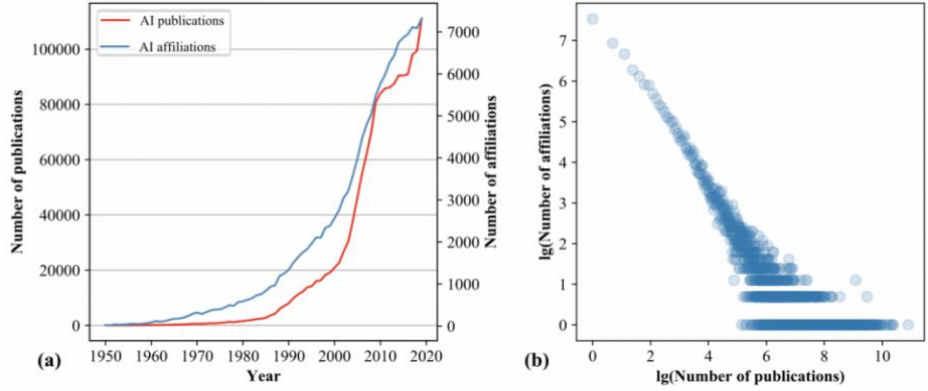

**Fig. 1.** Overview of the dataset. (a) Distribution of AI publications (blue) and distribution of unique affiliations in AI (red) over years. (b) The relationship between the number of AI publications and the number of affiliations in the dataset.

## 2.2 Measuring geographic distance

We define coauthors in a publication as a team and represent an AI publication $K$ with $m$ unique affiliations as $K = \{M_{K_1}, M_{K_2}, \dots, M_{K_m}\}$, in which $M_{K_i} = \langle t_k, Af_{k_i}, lat_{k_i}, lon_{k_i}, C_{k_i} \rangle \, (i \in \{1,2,\dots,m\})$, $t_k$ is the publication year of $K$, $Af_{k_i}$ denotes the $i^{th}$ affiliation id, $lat_{k_i}$ and $lon_{k_i}$ represents the latitude and longitude of $Af_{k_i}$, and $C_{k_i}$ is the country of $Af_{k_i}$. Then, the affiliation pairs for $K$ is expressed as $C(m,2)_K = \{(M_{K_1}, M_{K_2}), (M_{K_1}, M_{K_3}), \dots, (M_{K_i}, M_{K_j}), \dots \}$ ( $i \in \{1,2,\dots,m\}, j \in \{1,2,\dots,m\}, i \neq j$ ). Based on $(lat_{k_i}, lon_{k_i}) \in M_{k_i} \, (i \in \{1,2,\dots,m\})$, we employ GeoPy (https://geopy.readthedocs.io/en/stable/) to calculate the geographic distance for pairs in $C_K(m,2)$ as $D_k = \{D_{k_1 k_2}, D_{k_1 k_3}, \dots, D_{k_i k_j}, \dots \}$ ( $i \in \{1,2,\dots,m\}, j \in \{1,2,\dots,m\}, i \neq j$ ), in which $D_{k_i k_j}$ means the distance between the $i^{th}$ and $j^{th}$ affiliations, and the number of elements in $D_k$ is $\frac{m!}{[2!(m-2)!]}$. Therefore, the geographic distance (GD) of a team is defined at three levels:

(1) the average geographic distance (AveGD), defined as the average distance between affiliation pairs within a team, as expressed by:

$$AveGD = \frac{\sum D_{k_i k_j}}{\frac{m!}{[2!(m-2)!]}}, \quad D_{k_i k_j} \in D_K, \, i \in \{1,2,\dots,m\}, j \in \{1,2,\dots,m\}, i \neq j \; (kilometers) \quad (1)$$

(2) The maximum geographic distance (MaxGD), defined as the maximum distance among all affiliations within a team, as calculated by:

$$MaxGD = Max\left(D_{k_i k_j}\right), \quad D_{k_i k_j} \in D_K, \, i \in \{1,2,\dots,m\}, j \in \{1,2,\dots,m\}, i \neq j \; (kilometers) \quad (2)$$

(3) The minimum geographic distance (MinGD), defined as the minimum distance among all affiliations within a team, as represented by:

$$MinGD = Min\left(D_{k_i k_j}\right), \quad D_{k_i k_j} \in D_K, \, i \in \{1,2,\dots,m\}, j \in \{1,2,\dots,m\}, i \neq j \; (kilometers) \quad (3)$$

We conducted the geographic distance analysis on team collaboration in AI at domestic and international levels. For an AI publication $k$ with $m$ authors, the country set of its authors $C_K = \{C_{k_1}, C_{k_2}, \ldots, C_{k_m}\}$, if $C_{k_1} = C_{k_2} = \cdots = C_{k_m}$, it's a domestic collaboration; or, it's an international collaboration.

## 3 Results

### 3.1 Overview of team collaboration in AI

Multiple-affiliation collaborations in AI have gradually gained popularity over time. Fig. 2 (a) shows that the annual number of single-affiliation and multiple-affiliation AI publications both exhibit a noticeable increase since the late-1980s and early-2000s, respectively. After the rapid rise in the 21st century, the annual number of single-affiliation AI publications start-ed to stabilize with little fluctuation (around 65,000 publications per year), while that of multiple-affiliation ones continuously exhibit an upward trend. Although the number of single-affiliation AI publications has always been greater than that of multiple-affiliation ones, the percentage of multiple-affiliation AI publications exhibits a clear increasing trend, from 3% in 1951 to 38.4% in 2019.

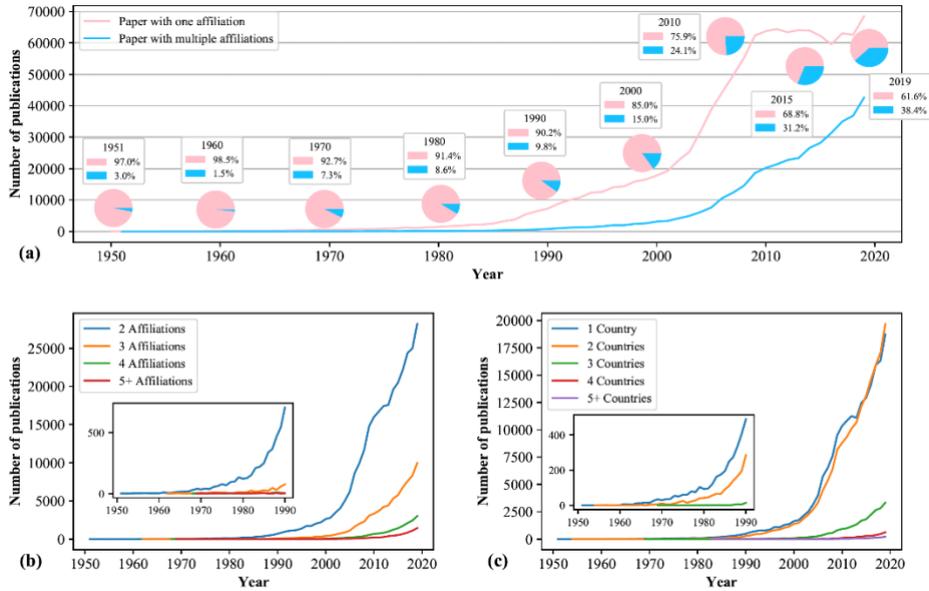

**Fig. 2.** Team collaboration in AI from the perspective of affiliations and countries during 1950-2019. (a) The changes in the number of publications with one affiliation (pink) and multiple affiliations (blue). The pies represent changes in the percentages of AI publications with one affiliation (pink) and multiple affiliations (blue). (b) The changes in the number of AI

publications collaborated by different number (2, 3, 4 and 5+) of affiliations. (c) The changes in the number of AI publications collaborated by different number (1,2,3,4 and 5+) of countries.

Fig. 2(b) and Fig. 2 (c) represent changes in the number of AI publications collaborated by a different number of affiliations and countries, respectively. Collabrations between two affiliations has been always the major form of cooperation in AI with a clear increasing trend since 1980. Collaborations among three or more affiliations also kept growing since 2000. During 1997-2000, around 25% AI publications were collaborated by 3 affiliations, and after 2010, nearly 50% AI papers were produced by 3+ affiliations. When considering the country distribution of collaborations in AI, most of AI publications were written by authors from a single country or two countries. Collaborations in AI among different numbers of countries all exhibit in-creasing trends. Before 2012, the number of two-country AI publications has always been slightly less than single-country ones; then, it surpassed the latter one and ended in the first place, indicating that international collaboration has become the main-stream in AI.

### 3.2 Geographic distance analysis

Overall, the geographic distances of team collaborations in AI exhibit an upward trend during 1950-2019 (Fig. 3). According to the growth rate of and the gaps among three geographic distances, we divided collaborations in AI into three stages, i.e., (1) stage 1(1950-1996): three geographic distances went up with large fluctuations, and there were no clear gaps among them; (2) stage 2 (1997-2009): the growth rate of three distances didn't change much, but the gaps among them were widening; (3) stage 3 (2010-2019): the growth rate of MaxGD and AveGD exhibit clear increasing trends again, and the gaps among them were further enlarged. To explicitly investigate team collaborations in AI, we divided team collaborations in AI into two categories: (1) domestic collaborations, in which authors of each publication are all from one country; and (2) inter-national collaborations, in which authors of each publications are from 2 or more countries.

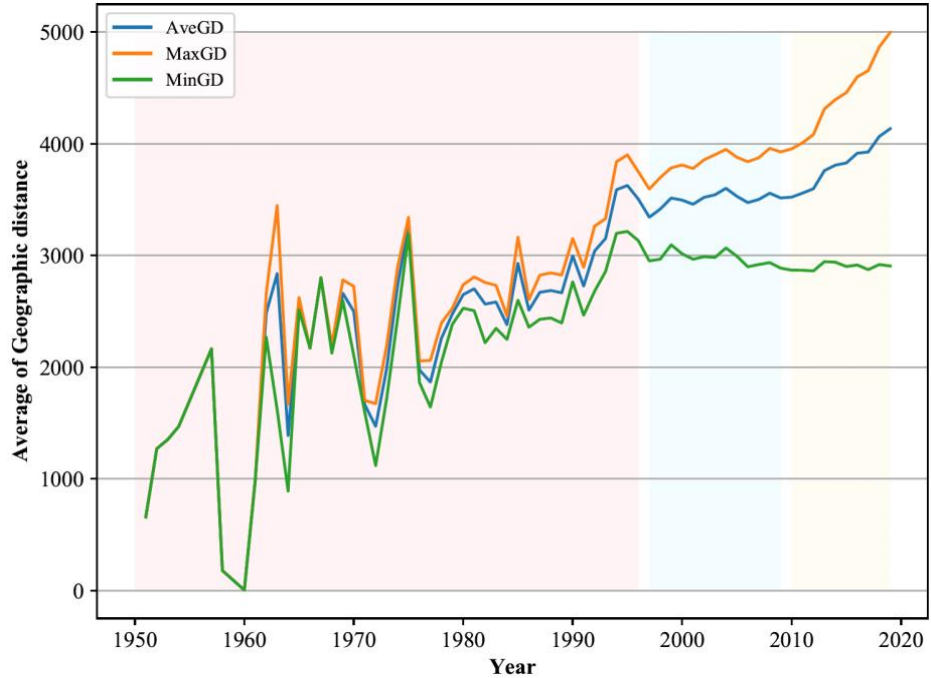

**Fig. 3.** The changes in the average values of AveGD, MaxGD and MinGD for all collaborations in AI. (Note, the background colors indicate the three stages of team collaborations in AI.)

(1) Domestic collaborations in AI

Fig. 4 shows the changes in the geographic distances for domestic collaborations in AI over time. In stage 1(1950-1996), the geographic distances of domestic collaborations in AI exhibit a decreasing trend, from around 1,000 kilometers in 1950 to 700 kilometers in 1996. The five most frequent domestic collaborations in AI all happened in the United states, and collaborations be-tween Harvard University and MIT ranked the first with 28 AI publications (Table 1). In stage 2 (1997-2009), the geographic distances of domestic collaborations steadily climbed to 800 kilometers. Although most of domestic collaborations were still occurred in the United States, Spain (the first) and China (the fifth) entered the top five list in this stage. In the last stage (2010-2019), the geographic distances of domestic collaborations swiftly increased, i.e., 950 kilometers for MinGD, 1200 kilometers for AveGD and 1,300 kilometers for MaxGD, respectively. Multiple research centers for AI has formed in the world, such as the United States and Singapore (Table 1). The swift growth of the MaxGD and MinGD both indicates that the breadth of domestic collaborations in AI has continuously increased.

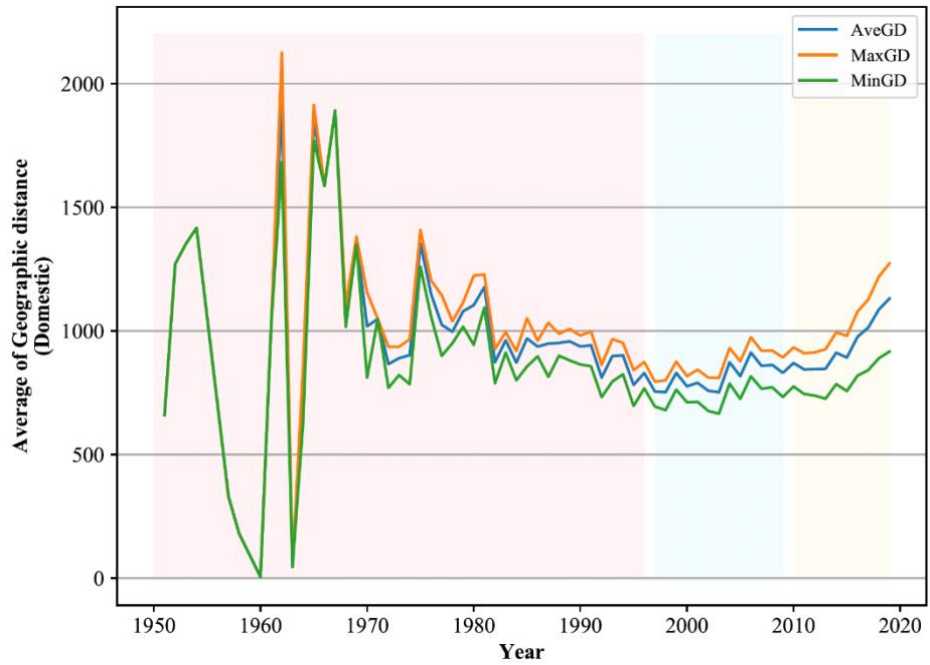

**Fig. 4.** The changes in the average values of AveGD, MaxGD and MinGD for domestic collaborations in AI (Note, the background colors indicate the three stages of team collaborations in AI.)

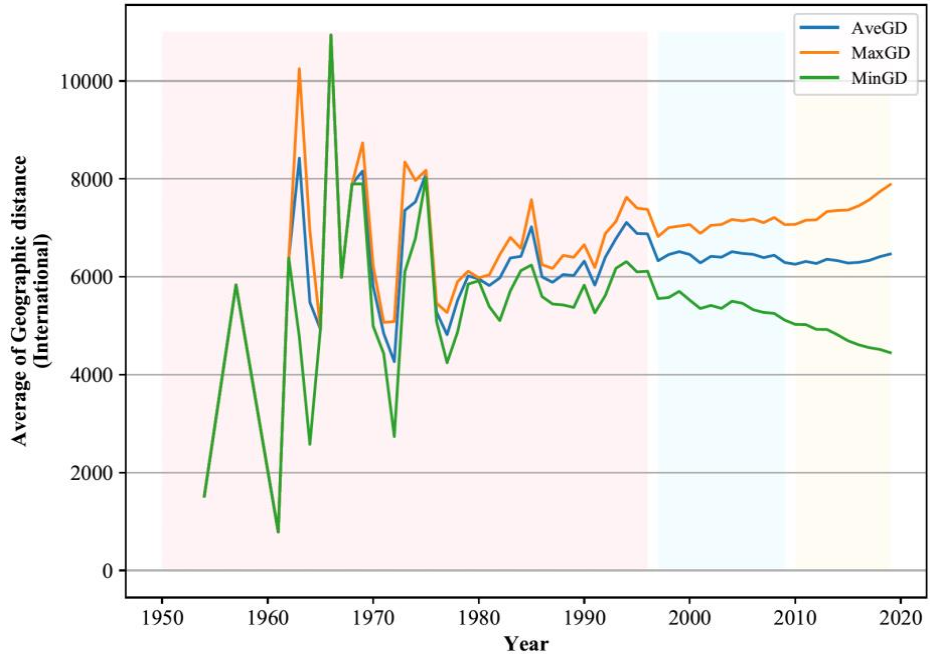

**Fig. 5.** The changes in the average values of AveGD, MaxGD and MinGD for international collaborations in AI. ((Note, the background colors indicate the three stages of team collaborations in AI.)

(2) International collaborations in AI

Fig. 5 shows the changes in the geographic distances for international collaborations in AI over time. Contrary to the domestic collaborations, in stage 1 (1950-1996), the geographic distances of international collaboration in AI exhibit an up-ward trend, from less 2,000 kilometers in 1950 to more than 6,000 kilometers in 1996. The most productive country in that peri-od was the United states followed by France and Germany; and the most frequent international partners in AI during 1950-1996 were Bell labs and Alcatel-Lucent, both of which have the strong back-ground of industry (Table 2). In stage 2 (1997-2009), the MaxGD and AveGD for international collaborations in AI kept still with small fluctuations, while the MinGD clearly declined to around 5,000 kilometers. The two most frequent international collaborations in AI were both between the United States and China, and Microsoft are the most popular affiliation internationally collaborated with affiliations in China. In stage 3 (2010-2019), the MaxGD of international collaborations in AI stably increased to 8,000 kilometers, while the MinGD swift decreased to around 4,200 Kilometers. This illustrates that both the breadth and density of international collaboration in AI have significantly increased recently. Universities in China participated in all the five most frequent international collaborations in AI, in which four were collaborations with Microsoft in the United States.

**Table 1.** Information about the 5 most frequent domestic collaborations in AI over three stages.

| Stages | Affiliation 1 | Affiliation 2 | Country | # of collaborations | Geographic distance(km) |
|---|---|---|---|---|---|
| Stage 1 (1950-1996) | Harvard Uni. | MIT | U.S. | 28 | 2.61 |
| | Northwestern Uni. | Uni. of Chicago | U.S. | 27 | 30.09 |
| | PARC | Stanford Uni. | U.S. | 24 | 3.31 |
| | Stanford Uni. | MIT | U.S. | 23 | 4,336.71 |
| | Stanford Uni. | Carnegie Mellon Uni. | U.S. | 20 | 3,641.54 |
| Stage 2 (1997-2009) | Uni. of Granada | Uni. of Jaen | Spain | 131 | 68.53 |
| | Carnegie Mellon Uni. | Uni. of Pittsburgh | U.S. | 121 | 0.83 |
| | MIT | Harvard Uni. | U.S. | 98 | 2.61 |
| | Uni. of Washington | Microsoft | U.S. | 86 | 12.93 |
| | National Taiwan Uni. | Academia Sinica | China | 86 | 8.65 |
| Stage 3 (2010-2019) | MIT | Harvard Uni. | U.S. | 273 | 2.61 |
| | Carnegie Mellon Uni. | Uni. of Pittsburgh | U.S. | 228 | 0.83 |
| | ASTR | Nanyang Technological Uni. | Singapore | 225 | 12.82 |
| | ASTR | National Uni. of Singapore | Singapore | 200 | 1.31 |
| | Islamic Azad Uni. | Amirkabir Uni. of Technology | Iran | 194 | 3.63 |

**Table 2.** Information on the 5 most frequent international collaborations in AI over three stages.

| Stages | Affiliation 1 | Affiliation 2 | Country pair | # of collaborations | Geographic distance (km) |
|---|---|---|---|---|---|
| Stage 1 (1950-1996) | Bell Labs | Alcatel-Lucent | (U.S., France) | 24 | 11,762.50 |
| | Karlsruhe Institute of Tech. | Indian Institute of Tech. Bombay | (Germany, India) | 16 | 6,573.54 |
| | AT&T | Alcatel-Lucent | (U.S., France) | 14 | 12,089.60 |
| | Ibaraki Uni. | Uni. of Alabama at Birmingham | (Japan, U.S.) | 11 | 10,847.97 |
| | Uni. of Manitoba | Hosei Uni. | (Canada, U.S.) | 11 | 9,010.36 |
| Stage 2 (1997-2009) | USTC | Microsoft | (China, U.S.) | 153 | 9,400.97 |
| | Tsinghua Uni. | Microsoft | (China, U.S.) | 131 | 8,714.75 |
| | Polish Academy of Sciences | Uni. of Alberta | (Poland, Canada) | 68 | 7,547.47 |
| | Hong Kong Uni. of Sci. and Tech. | Microsoft | (China, U.S.) | 64 | 10,434.14 |
| | Princeton Uni. | Siemens | (U.S., Germany) | 62 | 6,573.70 |

| Stage 3 (2010-2019) | USTC | Microsoft | (China, U.S.) | 367 | 9,400.97 |
| | Peking Uni. | Microsoft | (China, U.S.) | 242 | 8,716.74 |
| | Tsinghua Uni. | Microsoft | (China, U.S.) | 227 | 8,714.75 |
| | Chinese Academy of Sciences | Uni. of Tech., Sydney | (China, Australia) | 155 | 15,831.42 |
| | Harbin Institute of Tech. | Microsoft | (China, U.S.) | 140 | 12,214.72 |

## 4    Conclusion

From geographic distance perspective, a clear understanding of team collaboration patterns in AI during 1950-2019 was developed in this study. Three geographic distance indicators (AveGD, MaxGD and MinGD) based on the latitudes and longitudes of affiliations were employed, to conduct the analysis on team collaboration in AI at domestic and international levels.

We found that the amount of team collaborations in AI presented a solid growth over time. Overall, these collaborations can be divided into three stages according to the growth rates of and gaps among the three kinds of geographic distances. In stage 1 (1950-1996), the decline in the distances among domestic collaborations and the increase in distances among international collaborations indicates the growth of the density for domestic collaborations and the breadth for international collaborations. In stage 2 (1997-2009), although the distances for both were slightly changed, the frequency of collaborations evidently increased by the number of publications. In stage 3 (2010-2019), the breadth for both domestic collaborations further increased by the increase of MaxGD and AveGD, and the density of international collaborations showed a solid growth by the decline of MinGD. The United States produced the largest number of single-country and internationally collaborated AI publications over three stages, and China has played an important role in international collaborations in AI after 2010. In addition, industrial companies with ability of high-performance computing and massive data, such as Microsoft and Bell, have evident positive effect on the collaborations in AI.

For the future study, we will further measure the cultural distance and socioeconomic distance of team collaborations in AI. We will also investigate how these distances affect citation counts of AI publications, AI team diversity, the success of collaborations and scientific careers.

**Acknowledgement.** This work was supported by the National Science Foundation of China (71420107026).

# References

1. Yao, X., Zhang, C., Qu, Z., & Tan, B. C. Y.: Global village or virtual balkans? Evolution and performance of scientific collaboration in the information age. Journal of the Association for Information Science and Technology, 71(4), 395–408 (2020).

2. Bullock, J., Pham, K. H., Lam, C. S. N., & Luengo-Oroz, M.: Mapping the Landscape of Artificial Intelligence Applications against COVID-19. arXiv preprint arXiv:2003.11336 (2020).

3. Ai, T., Yang, Z., Hou, et al.: Correlation of Chest CT and RT-PCR Testing in Coronavirus Disease 2019 (COVID-19) in China: A Report of 1014 Cases. Radiology, 200642 (2020).

4. Hoffmann, M., Kleine-Weber, H., Schroeder, et al.: SARS-CoV-2 Cell Entry Depends on ACE2 and TMPRSS2 and Is Blocked by a Clinically Proven Protease Inhibitor. Cell, S0092867420302294 (2020).

5. Al-qaness, M. A. A., Ewees, A. A., Fan, H., & Abd El Aziz, M.: Optimization Method for Forecasting Confirmed Cases of COVID-19 in China. Journal of Clinical Medicine, 9(3), 674 (2020).

6. Niu, J., Tang, W., Xu, F., Zhou, X., & Song, Y.: Global research on artificial intelligence from 1990–2014: Spatially-explicit bibliometric analysis. ISPRS International Journal of Geo-Information, 5(5), 66 (2016).

7. Parreira, M. R., Machado, K. B., Logares, R., Diniz-Filho, J. A. F., & Nabout, J. C.: The roles of geographic distance and socioeconomic factors on international collaboration among ecologists. Scientometrics, 113(3), 1539–1550 (2017).

8. Fung, H. N., & Wong, C. Y.: Scientific collaboration in indigenous knowledge in context: Insights from publication and co-publication network analysis. Technological Forecasting and Social Change, 117, 57-69 (2017).

9. Yozwiak, N. L., Happi, C. T., Grant, D. S., Schieffelin, J. S., Garry, R. F., Sabeti, P. C., & Andersen, K. G.: Roots, not parachutes research collaborations combat outbreaks. Cell, 166(1), 5-8 (2016).

10. Sidone, O. J. G., Haddad, E. A., & Mena-Chalco, J. P.: Scholarly publication and collaboration in Brazil: The role of geography. Journal of the Association for Information Science and Technology, 68(1), 243–258 (2017).

11. Jiang, L., Zhu, N., Yang, Z., Xu, S., & Jun, M.: The relationships between distance factors and international collaborative research outcomes: A bibliometric examination. Journal of Informetrics, 12(3), 618–630 (2018).

12. Frank, M. R., Wang, D., Cebrian, M., & Rahwan, I.: The evolution of citation graphs in artificial intelligence research. Nature Machine Intelligence, 1(2), 79–85 (2019).

13. Tang, X., Li, X., Ding, Y., Song, M., & Bu, Y.: The pace of artificial intelligence innovations: Speed, talent, and trial-and-error. Journal of Informetrics, 14(4), 101094 (2020).